\documentclass{article} 

\usepackage{cite}
\usepackage{fullpage}
\newcommand{\shortcite}{\cite}

\usepackage{times}
\usepackage{caption}
\usepackage{soul}
\usepackage{url}
\usepackage{hyperref}
\usepackage{graphicx}
\usepackage{amsmath}
\usepackage{booktabs,makecell}
\usepackage{algorithm}
\usepackage{algorithmic}
\usepackage{subfigure}
\usepackage{booktabs}
\usepackage{amsmath,amssymb,amsthm,mathtools}
\usepackage{dblfloatfix}
\usepackage[usenames,dvipsnames]{xcolor}
\usepackage{array,multirow,bigdelim}
\usepackage{comment}
\usepackage{tikz}
\usepackage{pgfplots}
\usepackage{pgfplotstable}
\usetikzlibrary{positioning,intersections,calc,arrows}
\pgfplotsset{compat=newest}
\usepackage[colorinlistoftodos]{todonotes}
\usepackage[algo2e,ruled,vlined,linesnumbered]{algorithm2e}
\usepackage{capt-of}

\newcommand{\ot}{\leftarrow}

\newcommand{\argmin}{\mathop{\rm argmin}}
\renewcommand{\mid}{\,:\,}

\newcommand{\cW}{\mathcal{W}}
\newcommand{\cQ}{\mathcal{Q}}
\newcommand{\cL}{\mathcal{L}}
\newcommand{\cC}{\mathcal{C}}
\newcommand{\MV}{\mathtt{MV}}
\newcommand{\cE}{\mathcal{E}}

\newcommand{\pex}{p_{\mathrm{ex}}}

\newcommand{\nex}{n_{\mathrm{ex}}}
\newcommand{\nnon}{n_{\mathrm{non}}}

\theoremstyle{plain}
\newtheorem{theorem}     {Theorem}

\title{Graph Mining Meets Crowdsourcing: Extracting Experts \mbox{for Answer Aggregation}}

\author{{\bf Yasushi Kawase}\\
  {\normalsize Tokyo Institute of Technology and RIKEN AIP}\\
  {\normalsize \texttt{kawase.y.ab@m.titech.ac.jp}}
  \and 
  {\bf Yuko Kuroki}\\
  {\normalsize The University of Tokyo and RIKEN AIP} \\
  {\normalsize \texttt{ykuroki@ms.k.u-tokyo.ac.jp}}
  \and 
  {\bf Atsushi Miyauchi}\\
  {\normalsize RIKEN AIP}\\
  {\normalsize \texttt{atsushi.miyauchi.hv@riken.jp}}
}
\date{\empty}

\begin{document}
\maketitle

\begin{abstract}
Aggregating responses from crowd workers is a fundamental task in the process of crowdsourcing. 
In cases where a few experts are overwhelmed by a large number of non-experts, most answer aggregation algorithms such as the majority voting fail to identify the correct answers. 
Therefore, it is crucial to extract reliable experts from the crowd workers. 
In this study, we introduce the notion of \emph{expert core}, which is a set of workers that is very unlikely to contain a non-expert. 
We design a graph-mining-based efficient algorithm that exactly computes the expert core. 
To answer the aggregation task, we propose two types of algorithms. 
The first one incorporates the expert core into existing answer aggregation algorithms such as the majority voting, 
whereas the second one utilizes information provided by the expert core extraction algorithm pertaining to the reliability of workers. 
We then give a theoretical justification for the first type of algorithm. 
Computational experiments using synthetic and real-world datasets 
demonstrate that our proposed answer aggregation algorithms outperform state-of-the-art algorithms. 
\end{abstract}

\section{Introduction}\label{sec:introduction}
Crowdsourcing, which has become popular in recent years, is a process that requires completion of specific tasks by crowd workers.
In crowdsourced single-answer multiple-choice questions, workers are asked to select one answer out of multiple candidates for each given question.
Such a scheme is used, for instance, in
annotating named entities for microblogs~\cite{Finin2010},
sentiment classification on political blogs~\cite{Hsueh2009},
and image tagging~\cite{Lin2015}.

For the purpose of quality control, crowdsourcing systems usually assign multiple workers to the same questions and then aggregate their answers using some rule or algorithm.
A critical fact here is that the workers often have different levels of expertise, skills, and motivation; 
some workers may guess the correct answers with their rich knowledge, while others may answer almost randomly to get a reward without any effort.
Let us call the workers of the former and latter type \emph{experts} and \emph{non-experts}, respectively.

When only a few experts are overwhelmed by a large number of non-experts, 
the most intuitive answer aggregation rule, \emph{majority voting}, often fails to acquire the correct answers. 
Consider a small example in Table~\ref{table:typical}. 
There are six workers $w_1,\dots,w_6$ assigned to eight questions $q_1,\dots,q_8$.
For each question, the candidate answers are A, B, C, D, and E.
Among these workers, $w_1$, $w_2$, and $w_3$ are experts who almost always give the correct answers,
while the others, $w_4$, $w_5$, and $w_6$, are non-experts who give random answers. 
Let us apply the majority voting to their answers. 
The majority answer for $q_1$ is D, which is the correct answer. 
However, the majority answer for $q_8$ is C, which is not the correct answer.
In addition, we need tie-breaking for $q_3$ because B and D get the same number of votes.
It is very likely that as the fraction of non-experts increases, the quality of the majority voting answers deteriorates.

\begin{table}[t]
\caption{A small example of crowdsourced single-answer multiple-choice questions.}\label{table:typical}
\centering
\setlength{\tabcolsep}{4pt}
{\renewcommand{\arraystretch}{1}
\scalebox{1}{
\begin{tabular}{rc||*{8}{c}}
&$\cW\backslash\cQ$&$q_1$&$q_2$&$q_3$&$q_4$&$q_5$&$q_6$&$q_7$&$q_8$\\\cline{2-10}\addlinespace[1pt]
&\lower3pt\hbox{\scriptsize\shortstack{correct\\answer}}
     &D&C&D&E&B&C&A&E\\\addlinespace[1pt]\cline{2-10}
\ldelim\{{3}{27pt}[Experts]&$w_1$&D&C&D&E&B&C&A&E\\
&$w_2$&D&C&B&E&B&C&A&C\\
&$w_3$&D&C&D&D&B&C&A&E\\[2pt]
\ldelim\{{3}{45pt}[Non-experts]&$w_4$&C&B&A&A&E&C&D&B\\
&$w_5$&A&B&E&A&B&E&E&C\\
&$w_6$&C&A&B&E&B&B&A&C
\end{tabular}}}
\end{table}

Various answer aggregation algorithms exist in addition to the majority voting (see Related Work). 
However, most of these algorithms implicitly strengthen the majority answers and therefore fail to provide the true answers when the majority answers are incorrect. 
To overcome this issue, Li et al.~\shortcite{LBK2017} recently proposed a sophisticated answer aggregation algorithm for such a hard situation. 
More specifically, they introduced the notion of \emph{hyper questions}, each of which is a set of single questions. 
Their algorithm applies the majority voting (or other existing answer aggregation algorithms) for the hyper questions 
and then decodes the results to votes on individual questions. 
Finally, it applies the majority voting again to obtain the final answers.
The results of their experiments demonstrate that their algorithm outperforms existing algorithms.

\subsection{Our Contribution}
In this study, we further investigate the above-mentioned hard situation. 
Our contribution can be summarized as follows: 
\begin{enumerate}
\leftskip=4pt
\itemsep=1pt
\parsep=1pt
\item We introduce a graph-mining-based efficient algorithm that accurately extracts the set of experts; 
\item We propose two types of answer aggregation algorithms based on the above experts extraction algorithm; 
\item We provide a theoretical justification of our proposed algorithms; 
\item We conduct thorough computational experiments using synthetic and real-world datasets to evaluate the performance of our proposed algorithms. 
\end{enumerate}

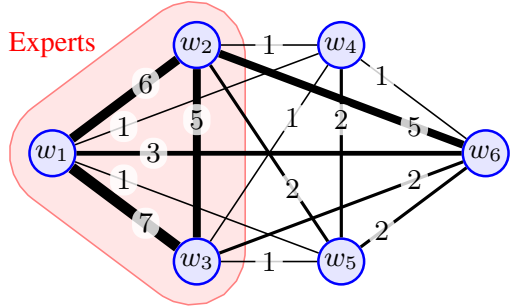
\begin{figure}[htbp]
\centering
\scalebox{1.2}{
\begin{tikzpicture}[thick,xscale=1.6,yscale=1.2,
   vertex/.style={circle,fill=blue!10,draw=blue,font=\small,inner sep=1pt},
   label/.style={font=\sffamily\small,rectangle,fill=white,opacity=0.8, text opacity=1,inner sep=2pt,rounded corners}]

 \draw[fill=red!10,draw=red!50,line width=31pt,rounded corners] (0,0) -- (1,1) -- (1,-1) -- cycle;
 \draw[fill=red!10,draw=red!10,line width=30pt,rounded corners] (0,0) -- (1,1) -- (1,-1) -- cycle;
 \node[text=red,font=\small] at (0,1) {Experts};

 \node[vertex] at (0,0) (1) {$w_1$};
 \node[vertex] at (1,1) (2) {$w_2$};
 \node[vertex] at (1,-1) (3) {$w_3$};
 \node[vertex] at (2,1) (4) {$w_4$};
 \node[vertex] at (2,-1) (5) {$w_5$};
 \node[vertex] at (3,0) (6) {$w_6$};

 \draw[line width=3.0pt] (1) -- (2) node[pos=.7,label] {$6$};
 \draw[line width=3.5pt] (1) -- (3) node[pos=.7,label] {$7$};
 \draw[line width=0.5pt] (1) -- (4) node[pos=.2,label] {$1$};
 \draw[line width=0.5pt] (1) -- (5) node[pos=.2,label] {$1$};
 \draw[line width=1.5pt] (1) -- (6) node[pos=.2,label] {$3$};
 \draw[line width=2.5pt] (2) -- (3) node[pos=.3,label] {$5$};
 \draw[line width=0.5pt] (2) -- (4) node[pos=.5,label] {$1$};
 \draw[line width=1.0pt] (2) -- (5) node[pos=.7,label] {$2$};
 \draw[line width=2.5pt] (2) -- (6) node[pos=.8,label] {$5$};
 \draw[line width=0.5pt] (3) -- (4) node[pos=.7,label] {$1$};
 \draw[line width=0.5pt] (3) -- (5) node[pos=.5,label] {$1$};
 \draw[line width=1.0pt] (3) -- (6) node[pos=.8,label] {$2$};
 \draw[line width=1.0pt] (4) -- (5) node[pos=.3,label] {$2$};
 \draw[line width=0.5pt] (4) -- (6) node[pos=.2,label] {$1$};
 \draw[line width=1.0pt] (5) -- (6) node[pos=.2,label] {$2$};
\end{tikzpicture}
}
\caption{An edge-weighted graph constructed from the example in Table~\ref{table:typical}.} \label{fig:network}
\end{figure}

\paragraph{First result.}
To design a powerful answer aggregation algorithm for the above hard situation, it is crucial to extract reliable experts from the crowd workers. 
In the example above, if we recognize that $w_1$, $w_2$, and $w_3$ are experts,
we can obtain the correct answers for all questions by simply applying the majority voting to the answers of the experts. 
The fundamental observation we use is as follows: as the experts almost always give the correct answers, each pair of experts frequently gives the same answer to a question.
Let us now consider constructing an edge-weighted complete undirected graph in which each vertex corresponds to a worker 
and each edge weight represents the agreement rate of the answers of two workers.
From the above observation, it is very likely that there is a dense component consisting of the experts, which we call the \emph{expert core}. 
Note that the formal definition of the expert core will be given in Section~\ref{sec:model}. 
Figure~\ref{fig:network} depicts an edge-weighted graph constructed from the example in Table~\ref{table:typical}.
As can be seen, experts $w_1$, $w_2$, and $w_3$ form a dense component.
Although, in this example, we simply set the edge weight to the number of same answers of two workers, we will use more suitable values in our algorithm.
To extract the expert core, we use a well-known dense subgraph extraction algorithm called the \emph{peeling algorithm}~\cite{Asahiro+_00}, which removes the most unreliable worker one by one.

\paragraph{Second result.}
Based on the expert core extraction algorithm, we propose two types of answer aggregation algorithms. 
The first one incorporates the expert core into existing answer aggregation algorithms such as the majority voting. 
Indeed, once we extract the expert core, we can apply existing algorithms to the answers of the workers in the expert core. 
The second one utilizes the information pertaining to the reliability of workers provided by the expert core extraction algorithm, 
which is quite effective when the task is very hard. 

\paragraph{Third result.}
We first demonstrate that the expert core is very unlikely to contain a non-expert if the number of questions is sufficiently large (Theorem~\ref{thm:asymp_excore}).
We then prove that the majority voting is asymptotically correct if there are only experts (Theorem~\ref{thm:mv_good}),
but not reliable if there are a large number of non-experts (Theorem~\ref{thm:mv_bad}).
Theorems~\ref{thm:asymp_excore} and \ref{thm:mv_good} provide a theoretical justification for the first type of our algorithm. 
In fact, combining these two theorems, we see that 
if the number of questions and the number of workers in the expert core are sufficiently large, 
our proposed algorithm (i.e., the majority voting among the workers in the expert core)
gives the correct answer with high probability. 
On the other hand, Theorem~\ref{thm:mv_bad} provides the limitation of the majority voting, 
which implies that it is quite important to exclude non-experts when we use the majority voting.

\paragraph{Fourth result.}
We conduct thorough computational experiments using synthetic and real-world datasets to evaluate our proposed algorithms. 
To simulate the hard situation in this study, 
we use six datasets recently collected by Li et al.~\shortcite{LBK2017} as real-world datasets, 
all of which have difficult heterogeneous-answer questions that require specialized knowledge. 
We demonstrate that the expert core counterparts of existing answer aggregation algorithms perform much better than their original versions. 
Furthermore, we show that our novel algorithm based on the information of the reliability of workers outperforms the other algorithms particularly when the task is quite hard.

\subsection{Related Work}
To date, a large body of work has been devoted to developing algorithms that estimate the quality of workers~\cite{DS1979,Whitehill2009,WBBP2010,RYZ+2010,KOS2011,WJ2011,Bachrach2012,DDC2012,KG2012,LPI2012,LLO+2012,ZPBM2012,AYL+2014,LLG+2014,VGK+2014,FLO+2015,MLL+2015,LBK2017,ZLLSC2017,LBK2018}.
In most existing work, the quality of workers and correct answers are estimated by an iterative approach comprising the following two steps:
(i) infer the correct answers based on the quality of workers estimated and (ii) estimate the quality of workers based on the correct answers inferred. 
For example, Whitehill et al.~\shortcite{Whitehill2009} modeled each worker's ability and each task's difficulty, 
and then designed a probabilistic approach, which they called \emph{GLAD} (Generative model of Labels, Abilities, and Difficulties).

Finding a dense component in a graph is a fundamental and well-studied task in graph mining.
A typical application of dense subgraph extraction is to identify components 
that have some special role or possess important functions in the underlying system represented by a graph. 
Examples include communities or spam link farms extraction in Web graphs~\cite{Dourisboure+_07,Gibson+_05}, 
identification of molecular complexes in protein--protein interaction graphs~\cite{Bader_Hogue_03}, 
and expert team formation in collaboration graphs~\cite{Bonchi+_14,Tsourakakis+_13}, 
The peeling algorithm~\cite{Asahiro+_00} is known to be effective in various optimization problems for dense subgraph extraction 
(e.g., the densest subgraph problem and its variations~\cite{Andersen_Chellapilla_09,Charikar_00,Kawase_Miyauchi_17,Miyauchi_Kakimura_18,Khuller_Saha_09} as well as other related problems~\cite{Tsourakakis+_13,Miyauchi_Kawase_15}).

\section{Model}\label{sec:model}
An instance of our problem is a tuple $(\cW,\cQ,\cC,\cL)$, where each component is defined as follows: 
There is a finite set of workers $\cW=\{w_1,\ldots,w_n\}$ and a finite set of questions $\cQ=\{q_1,\ldots,q_m\}$.
Each question $q$ has a set of candidate answers $\cC_q=\{c^q_1,\dots,c^q_s\}$ ($s\geq 2$).
Suppose that, for each question $q$, worker $w$ answers $l_{wq}\in\cC_q$ and let $\cL=(l_{wq})_{w\in\cW,\,q\in\cQ}$.
Our task is to estimate the unknown correct answers to the questions.

Suppose that, among the workers $\cW=\{w_1,\dots,w_n\}$, there are $\nex$ experts $\cE~(\subseteq \cW)$ who give the correct answer with probability $\pex~(>1/s)$,
and the other $\nnon~(=n-\nex)$ workers are non-experts who give an answer independently and uniformly at random.
If an expert makes a mistake, she selects a wrong answer independently and uniformly at random.
Thus, for a question $q\in\cQ$ with a correct answer $a_q\in\cC_q$ and an answer $c\in\cC_q$, it holds that
\begin{align*}
  \Pr[l_{wq}=c]=
  \begin{cases}
  \pex&(c=a_q,~w\in\cE),\\
  \frac{1-\pex}{s-1}&(c\ne a_q,~w\in\cE),\\
  1/s&(w\in\cW\setminus\cE).
  \end{cases}
\end{align*}
It should be noted that, by showing the candidate answers in random order for each worker,
we can handle some biases (e.g., some non-expert workers always choose the first candidate of answers) using this model. 

Let us consider the probability that a pair of workers $u,v\in\cW$ $(u\ne v)$ gives the same answer for a question $q\in\cQ$.
If at least one of $u$ and $v$ is a non-expert, then we have \(\Pr[l_{uq}=l_{vq}]=1/s\).
On the other hand, if both workers are experts, then \(\Pr[l_{uq}=l_{vq}]=\frac{(\pex-1/s)^2}{1-1/s}+\frac{1}{s}\),
which is strictly larger than $1/s$.

For each pair of workers $u,v\in\cW$,
let $\tau(u,v)$ be the number of questions such that $u$ and $v$ give the same answer,
that is, $\tau(u,v)=|\{q\in \cQ\mid l_{uq}=l_{vq}\}|$.
Here, if $\Pr[l_{uq}=l_{vq}]=p$, then $u$ and $v$ give the same answers for at least $\tau(u,v)$ questions
with probability $\sum_{i=\tau(u,v)}^{m}\binom{m}{i} p^i(1-p)^{m-i}$.

For a given $p\in(0,1)$, a subset of workers $W\subseteq\cW$ is called a \emph{$\theta$-expert set} with respect to $p$ if
\begin{align*}
  \prod_{v\in W\setminus\{u\}}\sum_{i=\tau(u,v)}^{m}\binom{m}{i} p^i(1-p)^{m-i}\le\theta
\end{align*}
holds for all $u\in W$.
Intuitively, a $\theta$-expert set with small $\theta$ is a set of workers that is very unlikely to contain a non-expert.
Let $\theta(W)$ be the minimum threshold such that $W$ is a $\theta$-expert set, that is, 
\begin{align*}
  \theta(W)=\max_{u\in W}\prod_{v\in W\setminus\{u\}}\sum_{i=\tau(u,v)}^{m}\binom{m}{i} p^i(1-p)^{m-i}.
\end{align*}
Then, we employ $W\subseteq\cW$ that minimizes $\theta(W)$ as the estimated set of experts.
We refer to such a set as the \emph{expert core} (with respect to $p$). 
As will be shown in Theorem~\ref{thm:asymp_excore}, the expert core is very unlikely to contain a non-expert if the number of questions is sufficiently large.

\section{Algorithms}\label{sec:algorithm}
In this section, we design an algorithm to compute the expert core, and then propose two types of answer aggregation algorithms. 
Our expert core extraction algorithm first constructs an edge-weighted complete undirected graph that represents the similarity of the workers in terms of their answers. 
Then, it extracts a dense component in the graph using the peeling algorithm~\cite{Asahiro+_00}.

\subsection{Peeling Algorithm}
We first revisit the peeling algorithm.
The algorithm iteratively removes a vertex with the minimum weighted degree in the current graph until we are left with only one vertex. 
Let $G=(V,E,\omega)$ be an edge-weighted graph.
For $S\subseteq V$ and $v\in S$, 
let $d_S(v)$ denote the weighted degree of $v$ in the induced subgraph $G[S]$, 
that is, $d_S(v)=\sum_{e=\{u,v\}\in E:\,u\in S}\omega(e)$. 
Then, the procedure can be summarized in Algorithm~\ref{alg:peeling}. 
Note that the algorithm here returns $S_i\in \{S_1,\dots, S_{|V|}\}$ that maximizes $d_{S_i}(v_i)$, 
although there are other variants depending on the problem at hand~\cite{Charikar_00,Kawase_Miyauchi_17,Miyauchi_Kawase_15,Tsourakakis+_13}. 
Algorithm~\ref{alg:peeling} can be implemented to run in $O(|E|+|V|\log |V|)$ time.

\begin{algorithm2e}[t]
  \SetKwInOut{Input}{Input}\Input{ Edge-weighted graph $G=(V,E,\omega)$}
  \SetKwInOut{Output}{Output}\Output{ $S\subseteq V$ ($k$-core with maximum $k$)}  
  \caption{Peeling algorithm}\label{alg:peeling}
    $S_{|V|}\ot V$\;
    \For{$i\ot |V|,\dots,2$}{
      \mbox{Find $v_i\in \argmin_{v\in S_i} d_{S_i}(v)$ and $S_{i-1}\ot S_i\setminus\{v_i\}$\;}
    }
    \Return $S_i\in \{S_1,\dots, S_{|V|}\}$ that maximizes $d_{S_i}(v_i)$\;
\end{algorithm2e}

Algorithm~\ref{alg:peeling} indeed produces the \emph{$k$-core decomposition} of graphs. 
For $G=(V,E,\omega)$ and a positive real $k$, a subset $S\subseteq V$ is called a \emph{$k$-core} 
if $S$ is a maximal subset in which every vertex $v\in S$ has a weighted degree of at least $k$ in the induced subgraph $G[S]$. 
Note that a $k$-core is unique for a fixed $k$. 
The $k$-core decomposition reveals the hierarchical structure of $k$-cores in a graph 
and is particularly focused on finding the $k$-core with maximum $k$. 
Algorithm~\ref{alg:peeling} is suitable for this scenario; in fact, it is evident that the algorithm returns the $k$-core with maximum $k$.

\subsection{Expert Core Extraction Algorithm}
Here, we present an algorithm to compute the expert core. 
In particular, we explain the construction of an edge-weighted graph, 
which represents the similarity of the workers in terms of their answers. 

In our algorithm, we set $p$ to the average agreement probability, that is, 
\begin{align*}
  p=\frac{1}{m}\sum_{q\in\cQ}\textstyle\bigl|\bigl\{\{u,v\}\in \binom{\cW}{2}\mid l_{uq}=l_{vq}\bigr\}\bigr|\big/\binom{n}{2},
\end{align*}
where $\binom{\cW}{2}=\bigl\{\{u,v\}\mid u,v\in \cW,\ u\neq v\bigr\}$, 
and extract the expert core with respect to this $p$ via the peeling algorithm.
We construct a complete graph $(\cW,\binom{\cW}{2})$ with weight $\gamma(u,v)=-\log \sum_{i=\tau(u,v)}^{m}\binom{m}{i}p^i(1-p)^{m-i}$, where recall that $\tau(u,v)=|\{q\in \cQ\mid l_{uq}=l_{vq}\}|$.
Then, we compute the $k$-core with maximum $k$ for the edge-weighted graph $(\cW,\binom{\cW}{2},\gamma)$ using Algorithm~\ref{alg:peeling}.
As a result, we can obtain a set of workers $W\subseteq \cW$ such that the following value is maximized: 
\begin{align*}
  \min_{u\in W}\sum_{v\in W\setminus\{u\}}\gamma(u,v)
  &=\min_{u\in W} \left(-\log\prod_{v\in W\setminus\{u\}}\sum_{i=\tau(u,v)}^{m}\binom{m}{i}p^i(1-p)^{m-i}\right)
  =-\log\theta(W).
\end{align*}
As $-\log x$ is monotone decreasing, the obtained set minimizes $\theta(W)$ and hence is the expert core.

Note that we can construct the edge-weighted graph $(\cW,\binom{\cW}{2},\gamma)$ in $O(n^2m)$ time
and Algorithm~\ref{alg:peeling} for the graph runs in $O(\binom{n}{2}+n\log n)=O(n^2)$ time.
Thus, we have the following theorem.
\begin{theorem}
The expert core can be found in $O(n^2m)$ time.
\end{theorem}

\subsection{Answer Aggregation Algorithms}
Once we extract the expert core, we can apply existing answer aggregation algorithms (e.g., the majority voting) 
to the answers of the workers in the expert core, 
which implies the expert core counterparts of the existing algorithms. 
In addition, we propose a novel answer aggregation algorithm. 
The peeling algorithm produces the ordering that represents the reliability of the workers.
With this ordering, we can obtain a majority-voting-based algorithm as follows.
At the beginning, we obtain the ordering of workers using the peeling algorithm for $(\cW,\binom{\cW}{2},\gamma)$.
Then we pick the most reliable pair of workers, that is, the pair left until the second last round of the peeling algorithm.
For each question, if the two workers select the same answer, then we employ this answer.
If the two workers select different answers, then we add the next most reliable worker one by one according to the ordering until two of them give the same answer.
Our algorithm, which we call \textsf{Top-2}, is summarized in Algorithm~\ref{alg:top2}. 
Note that the algorithm runs in $O(n^2m)$ time, which is the same as that of the expert core extraction algorithm.

\begin{algorithm2e}[t]
  \SetKwInOut{Input}{Input}\Input{ Worker set $\cW$; Question set $\cQ$; Answer set $\cL$}
  \SetKwInOut{Output}{Output}\Output{ Estimated true answers $(z_q)_{q\in\cQ}$}
  \SetInd{0.4em}{0.8em}
  \caption{\textsf{Top-2}}\label{alg:top2}
  Calculate the average agreement probability $p$\;
  Construct the edge-weighted graph $(\cW,\binom{\cW}{2},\gamma)$\;
  Let $S_1,\dots,S_n$ be the sets computed in Algorithm~\ref{alg:peeling}\;
  \For{$q\in\cQ$}{
    \For{$i\ot 2,\dots,n$}{
      \lIf{$\exists c^*\in\cC_q$ s.t. $|\{w\in W_{i}\mid l_{wq}=c^*\}|=2$}{
        $z_q\ot c^*$ and \textbf{break}
      }
      \lElseIf{$i=n$}{
        $z_q\ot l_{wq}$, where $\{w\}=S_1$
      }
    }
  }
  \Return $(z_q)_{q\in\cQ}$\;
\end{algorithm2e}

\section{Theoretical Results}\label{sec:theoretical}
In this section, we provide a theoretical justification for the first type of our proposed algorithm. 
The proofs can be found in Appendix~\ref{sec:omitted}.
Suppose that each worker gives answers according to the model described in Section~\ref{sec:model}. 

The following theorem states that the expert core does not contain any non-expert with high probability 
if the number of questions is sufficiently large.
\begin{theorem}\label{thm:asymp_excore}
  Let $W^*\subseteq \cW$ be the expert core.
  If $\nex\ge 2$ and $m\ge \frac{2n^4\log\frac{n^2}{\epsilon}}{(\pex-1/s)^4}$ for $\epsilon>0$, then we have
  $\Pr[W^*\subseteq \cE]\ge 1-\epsilon$.
\end{theorem}

The next theorem states that the majority voting gives the correct answer with high probability
if the number of workers is sufficiently large and all of them are experts. 
Let $\MV_q$ be the output of the majority voting for question $q\in\cQ$.
\begin{theorem}\label{thm:mv_good}
  If $n=\nex\ge \frac{2\log\frac{2s}{\epsilon}}{(\pex-1/s)^2}$ for $\epsilon>0$, then we have
  $\Pr\left[\MV_q=a_q\right]\ge 1-\epsilon$ $(\forall q\in\cQ)$.
\end{theorem}

Finally, the following theorem states that,
the majority voting does not provide the correct answer with high probability if the number of non-experts is much larger than the number of experts. 
\begin{theorem}\label{thm:mv_bad}
  If $\nnon\ge \frac{\nex^2}{2\pi\epsilon^2}$ for $\epsilon>0$, then we have
  $\Pr\left[\MV_q=a_q\right]\le \frac{1}{2}+\epsilon$ $(\forall q\in\cQ)$.
\end{theorem}

It should be noted that when the number of candidate answers is $2$, 
even the random choice gives the correct answer with probability $1/2$. 
Thus, the above theorem indicates that the majority voting is no better than the random choice if the number of non-experts is much larger than the number of experts.

\section{Experiments}
In this section, we report the results of computational experiments. 
The objective of our experiments is to examine the performance of our proposed algorithms 
from various aspects using both synthetic and real-world datasets.
The main component of our experiments comprises comparing our proposed answer aggregation algorithms with the following three existing algorithms: 
\textsf{MV} (the majority voting), \textsf{GLAD}~\cite{Whitehill2009}, 
and \textsf{Hyper-MV}~\cite{LBK2017}. 
As our proposed algorithms, we employ the expert core counterparts of the above three algorithms, 
which we denote by \textsf{Ex-MV}, \textsf{Ex-GLAD}, and \textsf{Ex-Hyper-MV}, respectively, 
in addition to \textsf{Top-2}.

\subsection{Datasets}
As for synthetic datasets, we generate a variety of instances 
using the model described in Section~\ref{sec:model}. 
Recall the following five parameters: 
number of workers $n=|\cW|$, number of questions $m=|\cQ|$, number of candidate answers (for each question) $s$, 
number of ground-truth experts $\nex$, and the probability $\pex$ that an expert gives the correct answer. 
Throughout the experiments, we set $s=5$.

\begin{table}
\caption{Real datasets used in our experiments. The last column gives the best accuracy rate (i.e., the fraction of the number of correct answers) among the workers. }\label{tab:instance}
\centering
\scalebox{1}{
\renewcommand{\arraystretch}{1.0}
\begin{tabular}{l*{4}{@{\hspace{8mm}}r}}
\toprule
Dataset     & $m$ & $n$ & $s$ & \textsc{Best}\\ 
\midrule
{\em Chinese    } & 24        & 50        & 5   & 0.79 \\ 
{\em English    } & 30        & 63        & 5   & 0.70 \\ 
{\em IT         } & 25        & 36        & 4   & 0.84 \\ 
{\em Medicine   } & 36        & 45        & 4   & 0.92 \\ 
{\em Pok\'{e}mon} & 20        & 55        & 6   & 1.00 \\ 
{\em Science    } & 20        & 111       & 5   & 0.85 \\ 
\bottomrule
\end{tabular}
}
\end{table}

Table~\ref{tab:instance} summarizes six datasets that we use as real-world datasets.
They were recently collected by Li et al.~\shortcite{LBK2017} 
using Lancers, a commercial crowdsourcing platform in Japan.
Consistent with the hard situation being addressed here, 
these datasets have difficult heterogeneous-answer questions that require specialized knowledge. 
In fact, as shown later, the majority voting performs poorly on these datasets. 
Note that the classical datasets used in other previous work (e.g., \emph{Bluebird} and \emph{Dog} in image tagging~\cite{WBBP2010,ZPBM2012}, \emph{Web} in Web search relevance judging~\cite{ZPBM2012}, \emph{Price} in product price estimation~\cite{LIS2013}, and \emph{RTE} and \emph{Temp} in textual entailment recognition~\cite{Snow+_08}) 
usually have relatively easy homogeneous-answer questions, 
which are not within the scope of this study. 
During the collection of the above six datasets, all workers were asked to answer all questions. 
This may not be the usual assumption in crowdsourcing but is effective in the current hard situation. 
Indeed, if we can identify reliable experts by asking all workers to answer a small number of questions at an early stage, 
then it is possible to ask only the identified experts to answer to the remaining questions, 
which may reduce the overall cost.
This scenario has been studied in previous work~\cite{LBK2017,LZF2014}.

\subsection{Experts Extraction}
To evaluate the performance of our expert core extraction algorithm in terms of extracting reliable experts,
we conducted simulations using synthetic datasets.
We set the number of workers as $n$ to $20$ and the number of experts as $\nex$ to $4$.
In these settings, we generated two types of instances to simulate different situations.
In the first type, to investigate the effect of the number of questions, 
we vary the number of questions $m$ from 5 to 50 (in increments of 5) 
under a fixed value of $\pex=0.8$.  
In the second type, to investigate the effect of the expertise of workers,
we vary the probability $\pex$ from 0.5 to 1.0 (in increments of 0.05) 
under a fixed value of $m=25$.  

As the sets of workers extracted by our peeling algorithm, 
we employ the Top-2 pair 
(i.e., a pair of workers left until the second last round of the peeling algorithm) 
and the expert core. 
To evaluate the quality of the set of workers at hand, we adopt the following two measures:
the \emph{precision} (number of ground-truth experts in the set divided by the size of the set) 
and the \emph{recall} (number of ground-truth experts in the set divided by the number of all ground-truth experts). 

\begin{figure}[t]
  \centering
  \includegraphics[scale=.82]{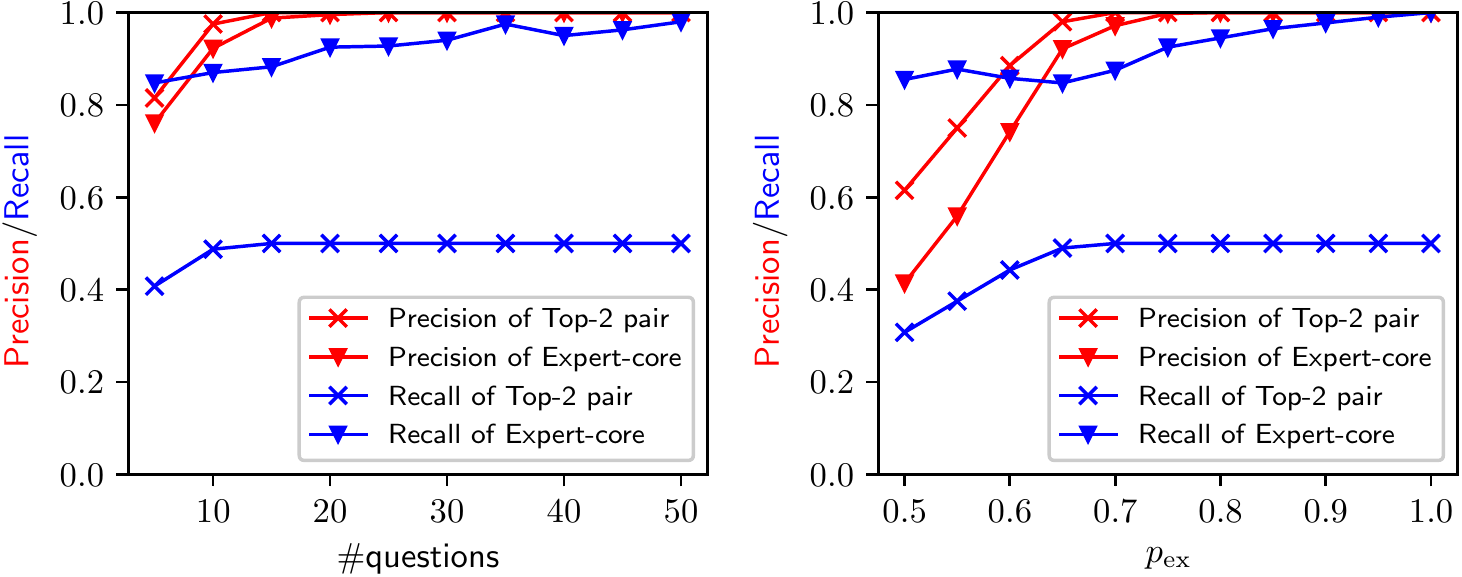}
  \caption{Results of experts extraction for synthetic datasets, where $n=20$ and $\nex=4$.}\label{fig:parameta}
\end{figure}

The results are shown in Figure~\ref{fig:parameta}.
Each point corresponds to an average over 100 data realizations.
As can be seen, the precision of the expert core becomes better as the number of questions $m$ or the probability $\pex$ increases.
Moreover, the expert core is robust in terms of the recall; 
in fact, even when $\pex$ is small (i.e., $\pex \leq 0.7$), 
the expert core achieves an average recall of $0.8$. 
The Top-2 pair achieves better precision in all parameter settings. 

\subsection{Ordering Defined by Peeling}
The objective here is to demonstrate that the ordering of workers 
defined by the peeling algorithm for $(\cW,\binom{\cW}{2},\gamma)$ reflects the accuracy rate of the workers. 
To this end, we perform the peeling algorithm for the real-world datasets. 
The results are shown in Figure~\ref{fig:rorder}. 
In the subfigures, the horizontal axis represents the ordering of workers defined by the peeling algorithm; 
specifically, the peeling algorithm has removed the workers from right to left. 
The vertical axis represents the accuracy rate of the workers. 
The bars corresponding to the workers contained in the expert core are colored black, 
while those corresponding to the other workers are colored gray.

\newcommand{\wid}{5cm}
\begin{figure}[t]
  \centering
  \subfigure{\label{fig:chinese_order}
    \includegraphics[width=\wid]{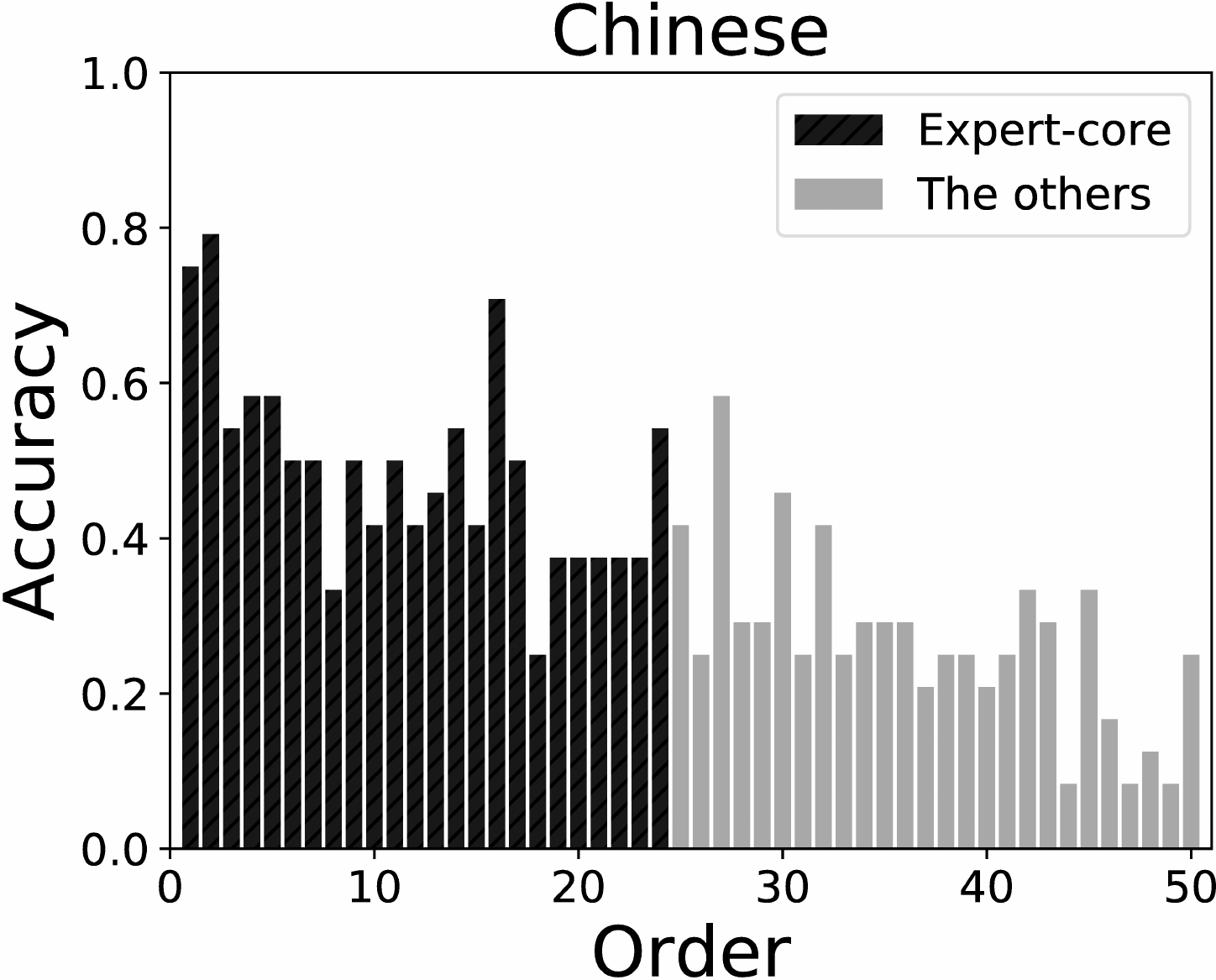}
  }
  \subfigure{\label{fig:english_order}
    \includegraphics[width=\wid]{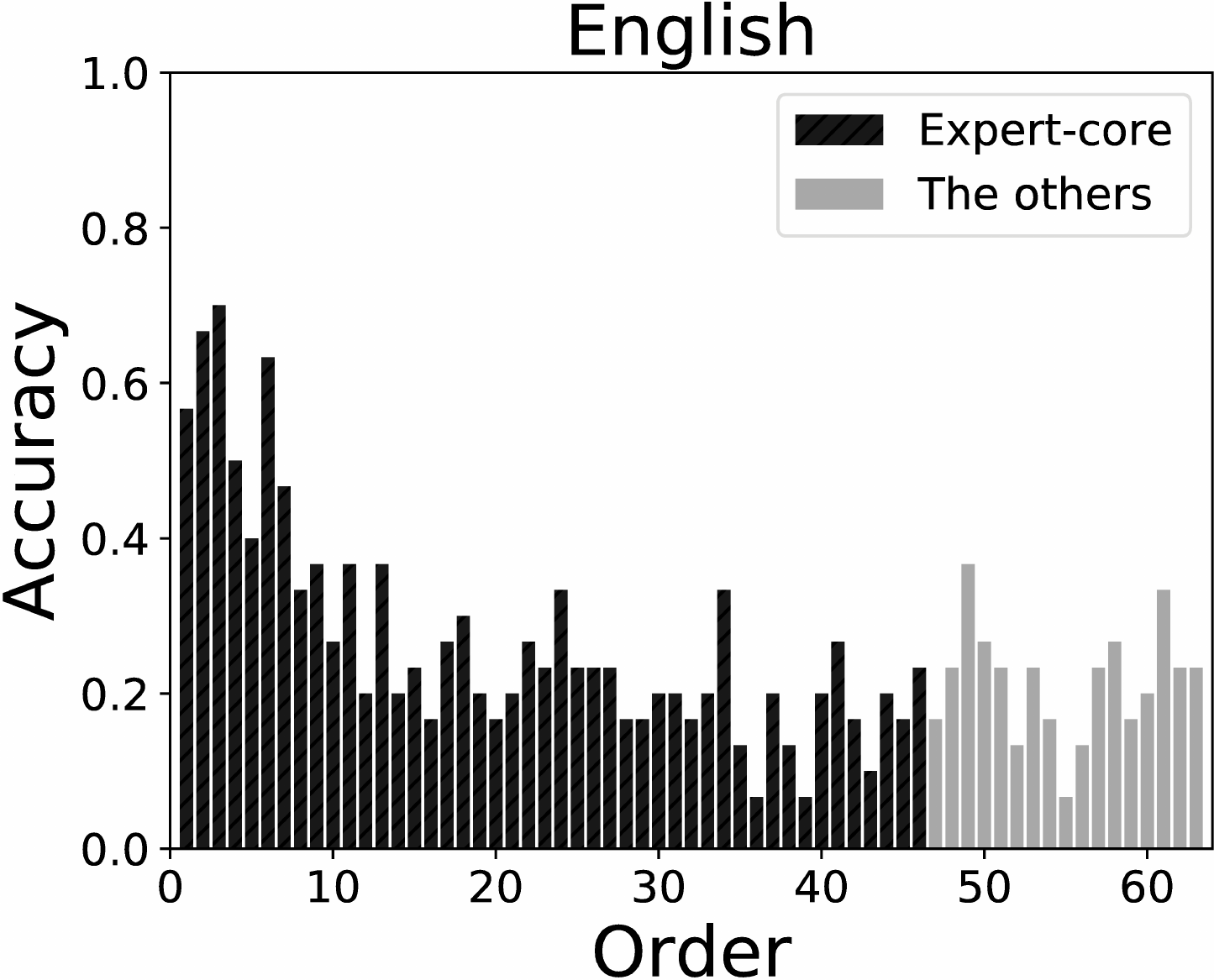}
  }
  \subfigure{\label{fig:it_order}
    \includegraphics[width=\wid]{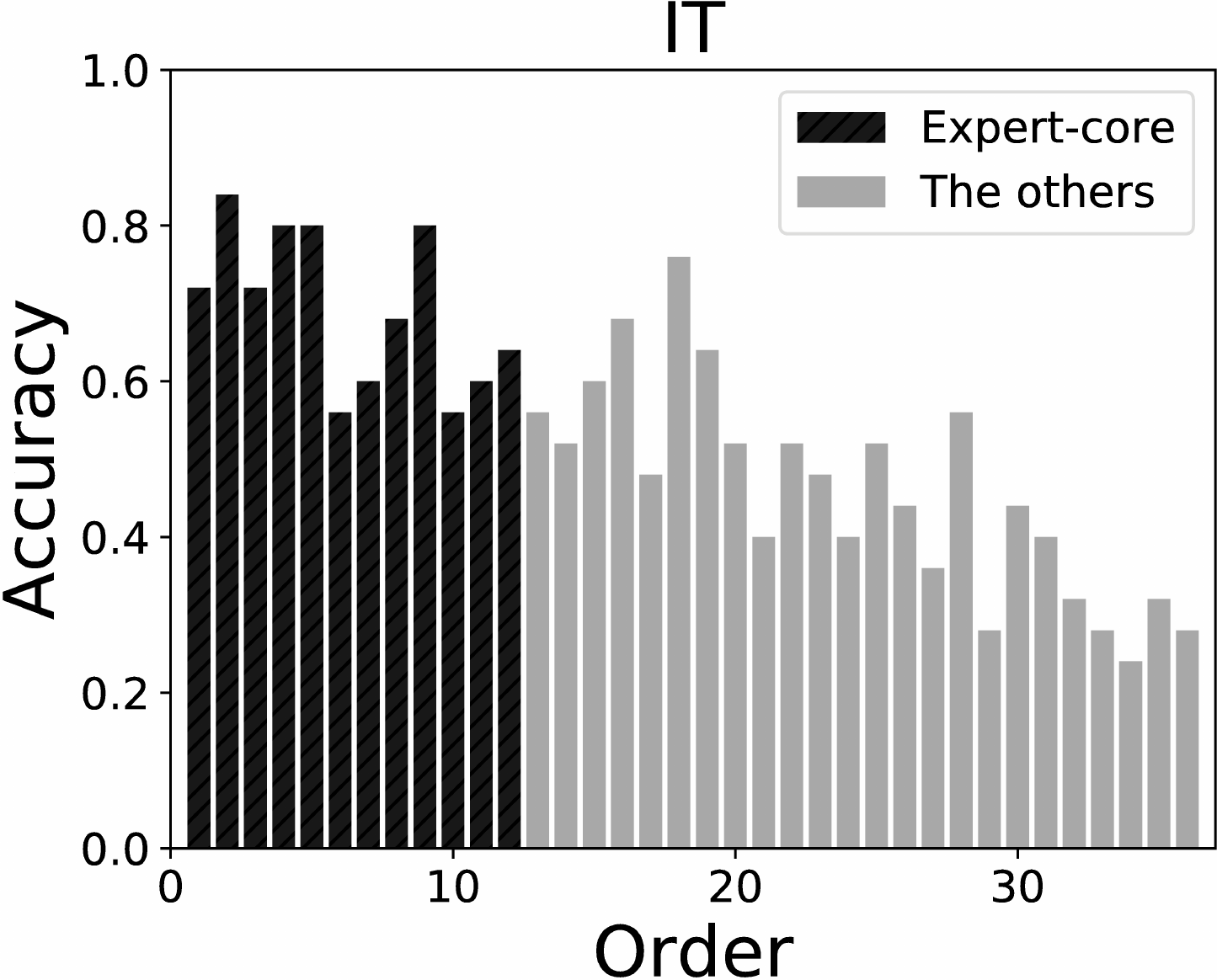}
  }
  \subfigure{\label{fig:medicine_order}
    \includegraphics[width=\wid]{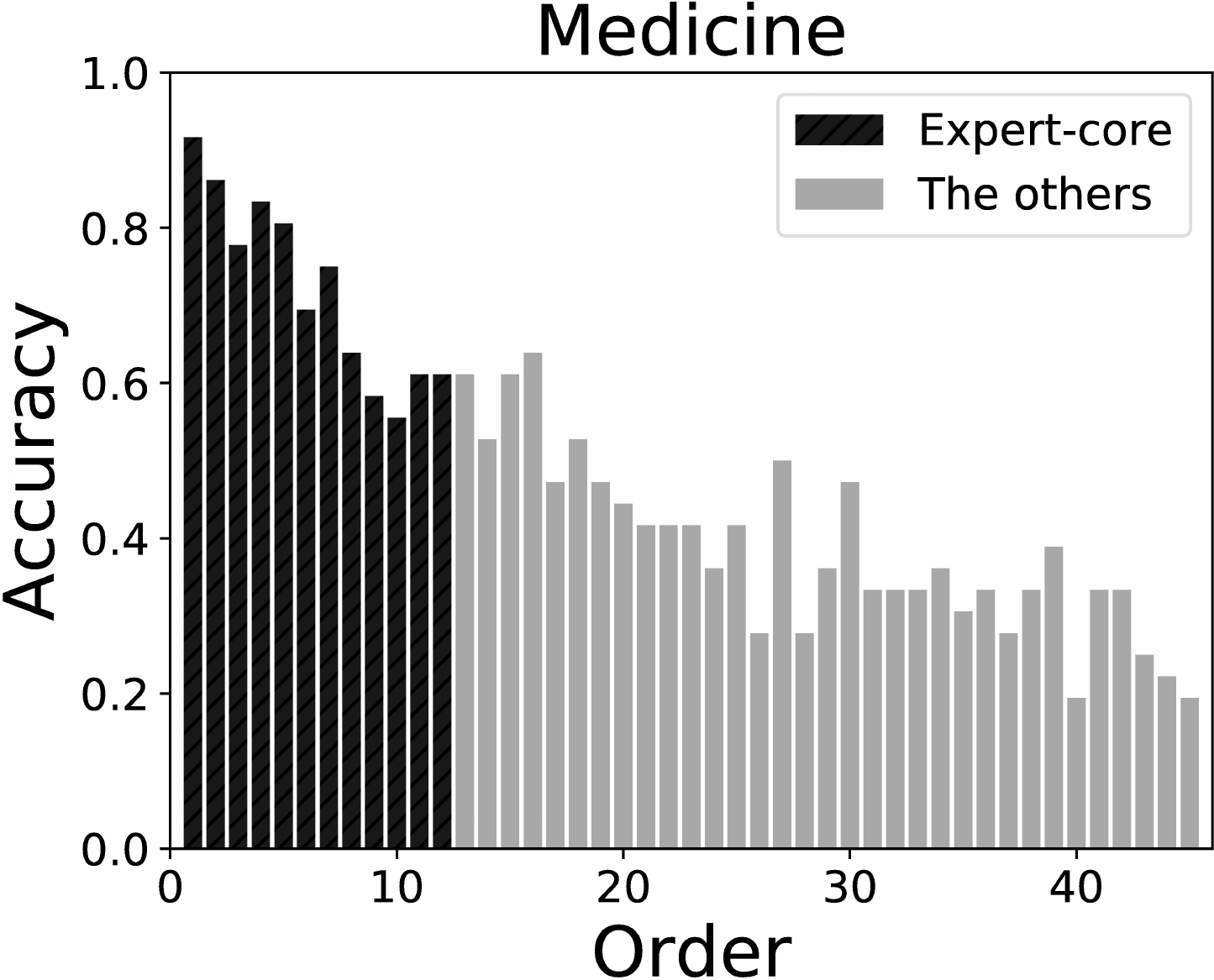}
  }
  \subfigure{\label{fig:pokemon_order}
    \includegraphics[width=\wid]{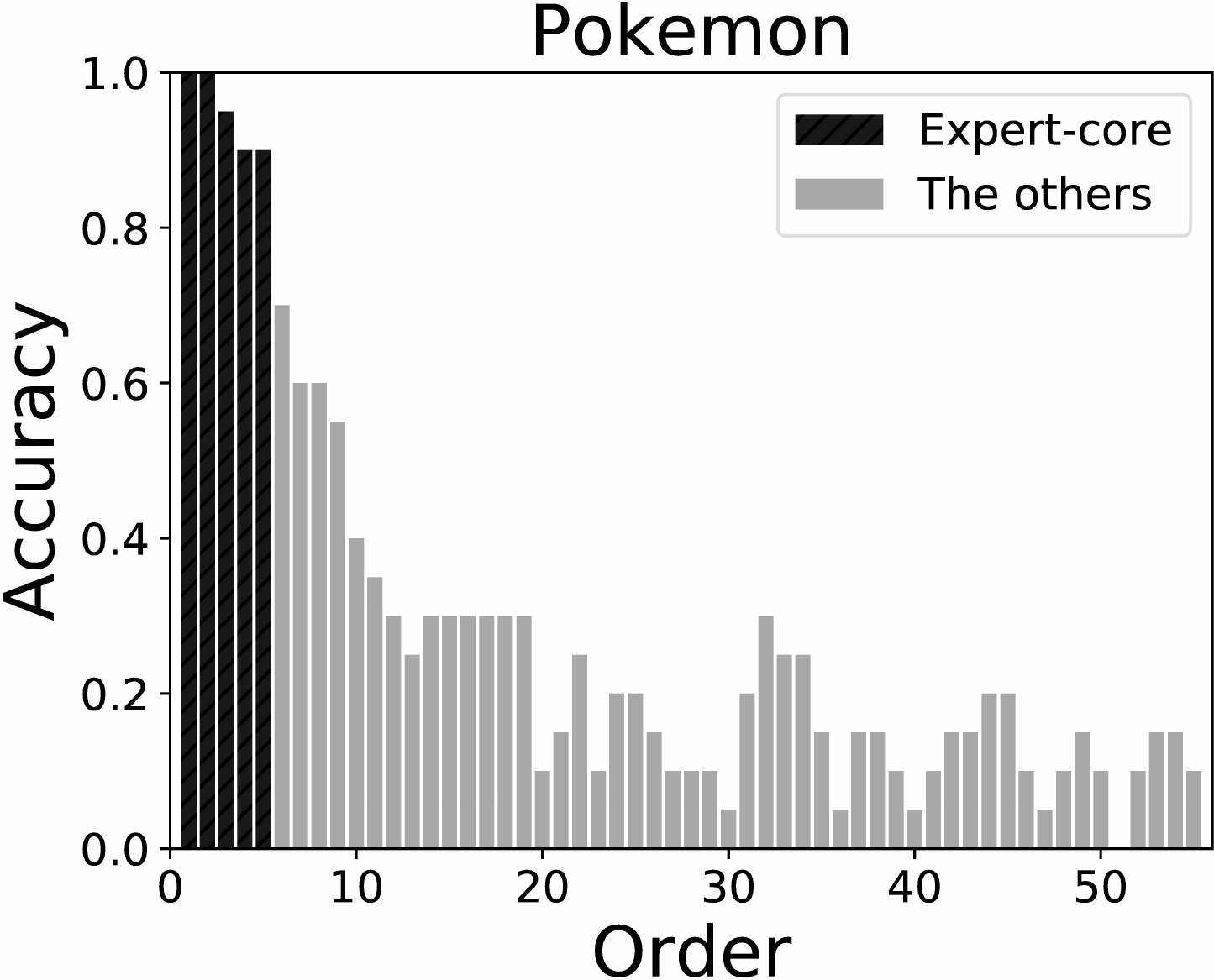}
  }
  \subfigure{\label{fig:science_order}
    \includegraphics[width=\wid]{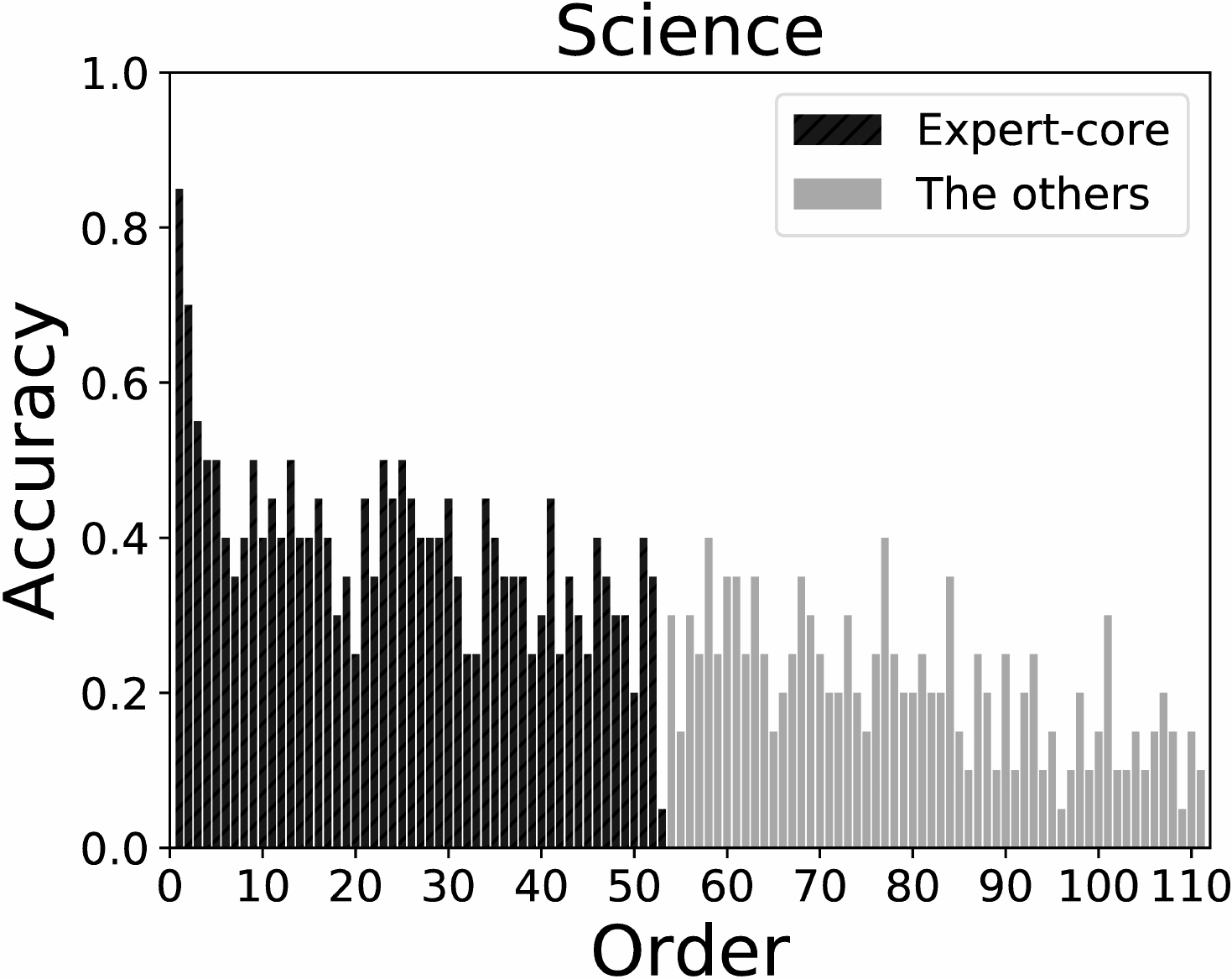}
  }   
\caption{Ordering defined by our peeling algorithm for real-world datasets.}\label{fig:rorder}
\end{figure}

As can be seen, workers with smaller order tend to have higher accuracy rates, 
In addition, the expert core contains almost all workers with significantly high accuracy rates.
which demonstrates the reliability of our peeling algorithm. 
In particular, for {\em Pok\'{e}mon} and {\em Science},
the ordering defined by the peeling algorithm correctly reproduces the ordering of the top five workers in terms of accuracy rates.

\begin{table*}[t!]
\caption{Results of answer aggregation algorithms for synthetic datasets, where $n=100$ and $m=50$. Each cell gives the average and standard deviation of the accuracy rate over 100 data realizations. For each setting, the best average accuracy rate 
is written in bold.}\label{table:synthetic}
\centering
{\renewcommand{\arraystretch}{1.0}
\tabcolsep=3.5mm
\scalebox{1}{
\begin{tabular}{cccccccc}
\toprule
$(\pex,\nex)$ & \textsf{Top-2} & \textsf{Ex-MV} & \textsf{Ex-GLAD} & \textsf{Ex-Hyper-MV} & \textsf{MV} & \textsf{GLAD} & \textsf{Hyper-MV}  \\ 
\midrule
\rule{0pt}{6mm} \raise1mm\hbox{$(0.7,  2)$}  & \shortstack{{ \bf 0.519 }\\ {\scriptsize$\pm$0.218}} & \shortstack{0.285\\ {\scriptsize$\pm$0.079}} & \shortstack{0.277\\ {\scriptsize$\pm$0.098}} & \shortstack{0.276\\ {\scriptsize$\pm$0.079}} & \shortstack{0.273\\ {\scriptsize$\pm$0.058}} & \shortstack{0.264\\ {\scriptsize$\pm$0.092}} & \shortstack{0.272\\ {\scriptsize$\pm$0.062}} \\ \Xhline{0.1pt}
\rule{0pt}{6mm} \raise1mm\hbox{$(0.7,  4)$}  & \shortstack{{ \bf 0.892 }\\ {\scriptsize$\pm$0.047}} & \shortstack{0.388\\ {\scriptsize$\pm$0.120}} & \shortstack{0.563\\ {\scriptsize$\pm$0.212}} & \shortstack{0.513\\ {\scriptsize$\pm$0.120}} & \shortstack{0.348\\ {\scriptsize$\pm$0.070}} & \shortstack{0.549\\ {\scriptsize$\pm$0.212}} & \shortstack{0.514\\ {\scriptsize$\pm$0.131}} \\ \Xhline{0.1pt}
\rule{0pt}{6mm} \raise1mm\hbox{$(0.7,  6)$}  & \shortstack{{ \bf 0.911 }\\ {\scriptsize$\pm$0.040}} & \shortstack{0.624\\ {\scriptsize$\pm$0.247}} & \shortstack{0.886\\ {\scriptsize$\pm$0.073}} & \shortstack{0.782\\ {\scriptsize$\pm$0.247}} & \shortstack{0.426\\ {\scriptsize$\pm$0.066}} & \shortstack{0.865\\ {\scriptsize$\pm$0.083}} & \shortstack{0.755\\ {\scriptsize$\pm$0.106}} \\ \Xhline{0.1pt}
\rule{0pt}{6mm} \raise1mm\hbox{$(0.8,  2)$}  & \shortstack{{ \bf 0.750 }\\ {\scriptsize$\pm$0.097}} & \shortstack{0.320\\ {\scriptsize$\pm$0.110}} & \shortstack{0.300\\ {\scriptsize$\pm$0.131}} & \shortstack{0.405\\ {\scriptsize$\pm$0.110}} & \shortstack{0.284\\ {\scriptsize$\pm$0.069}} & \shortstack{0.303\\ {\scriptsize$\pm$0.139}} & \shortstack{0.380\\ {\scriptsize$\pm$0.126}} \\ \Xhline{0.1pt}
\rule{0pt}{6mm} \raise1mm\hbox{$(0.8,  4)$}  & \shortstack{{ \bf 0.956 }\\ {\scriptsize$\pm$0.026}} & \shortstack{0.783\\ {\scriptsize$\pm$0.259}} & \shortstack{0.896\\ {\scriptsize$\pm$0.138}} & \shortstack{0.866\\ {\scriptsize$\pm$0.259}} & \shortstack{0.373\\ {\scriptsize$\pm$0.069}} & \shortstack{0.814\\ {\scriptsize$\pm$0.160}} & \shortstack{0.813\\ {\scriptsize$\pm$0.095}} \\ \Xhline{0.1pt}
\rule{0pt}{6mm} \raise1mm\hbox{$(0.8,  6)$}  & \shortstack{0.958\\ {\scriptsize$\pm$0.028}} & \shortstack{{ \bf 0.984 }\\ {\scriptsize$\pm$0.018}} & \shortstack{0.982\\ {\scriptsize$\pm$0.020}} & \shortstack{0.969\\ {\scriptsize$\pm$0.018}} & \shortstack{0.484\\ {\scriptsize$\pm$0.065}} & \shortstack{0.950\\ {\scriptsize$\pm$0.031}} & \shortstack{0.962\\ {\scriptsize$\pm$0.030}} \\ \Xhline{0.1pt}
\rule{0pt}{6mm} \raise1mm\hbox{$(0.9,  2)$}  & \shortstack{{ \bf 0.879 }\\ {\scriptsize$\pm$0.048}} & \shortstack{0.356\\ {\scriptsize$\pm$0.172}} & \shortstack{0.384\\ {\scriptsize$\pm$0.214}} & \shortstack{0.702\\ {\scriptsize$\pm$0.172}} & \shortstack{0.290\\ {\scriptsize$\pm$0.067}} & \shortstack{0.353\\ {\scriptsize$\pm$0.180}} & \shortstack{0.704\\ {\scriptsize$\pm$0.146}} \\ \Xhline{0.1pt}
\rule{0pt}{6mm} \raise1mm\hbox{$(0.9,  4)$}  & \shortstack{{ \bf 0.991 }\\ {\scriptsize$\pm$0.014}} & \shortstack{0.989\\ {\scriptsize$\pm$0.015}} & \shortstack{0.991\\ {\scriptsize$\pm$0.014}} & \shortstack{0.988\\ {\scriptsize$\pm$0.015}} & \shortstack{0.416\\ {\scriptsize$\pm$0.063}} & \shortstack{0.941\\ {\scriptsize$\pm$0.091}} & \shortstack{0.982\\ {\scriptsize$\pm$0.021}} \\ \Xhline{0.1pt}
\rule{0pt}{6mm} \raise1mm\hbox{$(0.9,  6)$}  & \shortstack{0.989\\ {\scriptsize$\pm$0.014}} & \shortstack{{ \bf 0.998 }\\ {\scriptsize$\pm$0.005}} & \shortstack{0.998\\ {\scriptsize$\pm$0.006}} & \shortstack{0.996\\ {\scriptsize$\pm$0.005}} & \shortstack{0.550\\ {\scriptsize$\pm$0.073}} & \shortstack{0.984\\ {\scriptsize$\pm$0.016}} & \shortstack{0.997\\ {\scriptsize$\pm$0.007}} \\
\bottomrule
\end{tabular}
}}
\vskip 16pt
\caption{Results of answer aggregation algorithms for real-world datasets. Each cell gives the accuracy rate. The average and standard deviation over 100 runs are listed for the hyper-question-based algorithms. For each dataset, the best accuracy rate is written in bold.}\label{table:real}
\centering
{\renewcommand{\arraystretch}{1.0}
\tabcolsep=3.5mm
\scalebox{1}{
\begin{tabular}{ccccccccc}
\toprule
Dataset &\textsf{Top-2}& \textsf{Ex-MV} & \textsf{Ex-GLAD} & \textsf{Ex-Hyper-MV} & \textsf{MV} & \textsf{GLAD}  & \textsf{Hyper-MV} \\
\midrule
Chinese &{\bf 0.750}  &0.667 & 0.667  & 0.700 {\footnotesize($\pm$0.026)}& 0.625 & 0.542  &  0.696 {\footnotesize($\pm$0.042)} \\ 
English&{\bf 0.733}&0.400 & 0.533  & 0.490 {\footnotesize($\pm$0.049)}     & 0.467 & 0.567  &  0.542 {\footnotesize($\pm$0.060)} \\ 
IT&0.800&0.760 & 0.800 & 0.802 {\footnotesize($\pm$0.008)}     & 0.760 & 0.720  & {\bf 0.828} {\footnotesize($\pm$0.018)} \\ 
Medicine&	0.944&{\bf 0.972 }& {\bf 0.972}  & 0.951 {\footnotesize($\pm$0.022)}    & 0.667 & 0.694  &  0.848 {\footnotesize($\pm$0.019)} \\ 
Pok\'{e}mon&{\bf 1.000}& {\bf 1.000} & {\bf 1.000}  & {\bf 1.000} {\footnotesize($\pm$0.000)}     & 0.650 & 0.850  & {\bf  1.000} {\footnotesize($\pm$0.000)} \\ 
Science&	{\bf 0.900} &0.650 & 0.650  & 0.603 {\footnotesize($\pm$0.025)}     & 0.550 & 0.550 &  0.606 {\footnotesize($\pm$0.018)} \\
\bottomrule
\end{tabular}
}}
\end{table*}

\subsection{Comparison with Existing Algorithms}
Using both synthetic and real-world datasets,
we compare the performance of our proposed answer aggregation algorithms, 
that is, \textsf{Top-2}, \textsf{Ex-MV}, \textsf{Ex-GLAD}, and \textsf{Ex-Hyper-MV}, 
with existing algorithms, that is, \textsf{MV}, \textsf{GLAD}, and \textsf{Hyper-MV}.
We first explain the details of \textsf{GLAD} and \textsf{Hyper-MV}.
\textsf{GLAD} is a method that takes into account not only the worker expertise, denoted by $\alpha$, 
but also the difficulty of each question, denoted by $\beta$, 
We set $\alpha \sim \mathcal{N}(1,1)$ and $\beta \sim \mathcal{N}(1,1)$ as in Li et al.~\shortcite{LBK2017}.
It is known that \textsf{GLAD} runs in $O(nmsT)$ time, where $T$ is the number of iterations to converge. 
\textsf{Hyper-MV} is a method that applies the majority voting 
to the hyper questions rather than the original individual questions.
Because the number of possible hyper questions may be too large, 
Li et al.~\shortcite{LBK2017} suggested to apply the following random sampling procedure for $r$ times: 
(i) shuffle the order of all single questions uniformly at random to generate a permutation and 
(ii) from this permutation, pick every $k$ single questions from the beginning of the queue to generate hyper questions as long as they can be picked. 
Then, the overall time complexity of \textsf{Hyper-MV} is given by $O(rmn)$. 
We performed the sampling procedure with $k=5$ for $r=100$ times, as suggested by Li et al.~\shortcite{LBK2017}.

Table~\ref{table:synthetic} shows the results for synthetic datasets.
We list the results for nine settings 
in which the probability $\pex$ varies in $\{0.7, 0.8, 0.9\}$ and $\nex$ varies in $\{2, 4, 6\}$.
We set the number of workers $n$ to 100 and the number of questions $m$ to 50.
As can be seen, each expert core counterpart achieves better performance than the original algorithm. 
In particular, \textsf{Ex-MV} significantly improves the performance of \textsf{MV}. 
\textsf{Top-2} outperforms the other algorithms particularly when the problem is quite difficult, 
although \textsf{Ex-MV} performs better when the problem is relatively easy. 

Table~\ref{table:real} summarizes the results for the six real-world datasets.
As can be seen, for all datasets except {\em IT}, our proposed algorithms achieve the best performance. 
In fact, for almost all datasets, the performance of the existing algorithms is improved by using the expert core.
Among our proposed algorithms, \textsf{Top-2} provides the best performance; 
in particular, for {\em English} and {\em Science}, the accuracy rate of \textsf{Top-2} is even higher than those of the other algorithms.
It should be noted that, for {\em English}, {\em Medicine}, and {\em Science}, 
the accuracy rate of \textsf{Top-2} is strictly higher than the best accuracy rate among workers (presented in Table~\ref{tab:instance}), 
which emphasizes the power of answer aggregation in crowdsourcing. 
According to the trend observed in synthetic datasets, 
it is very likely that the high performance of \textsf{Top-2} stems from the high fraction of non-experts in real-world datasets.

\section{Conclusion}

In this study, we addressed the answer aggregation task in crowdsourcing. 
Specifically, we focused on a hard situation wherein a few experts are overwhelmed by a large number of non-experts. 
To design powerful answer aggregation algorithms for such situations, 
we introduced the notion of \emph{expert core}, which is a set of workers that is very unlikely to contain a non-expert. 
We then designed a graph-mining-based efficient algorithm that exactly computes the expert core. 
Based on the expert core extraction algorithm, we proposed two types of answer aggregation algorithms. 
The first one incorporates the expert core into existing answer aggregation algorithms. 
The second one utilizes the information provided by the expert core extraction algorithm pertaining to the reliability of workers. 
In particular, we provided a theoretical justification for the first type of algorithm. 
if the number of questions and the number of workers in the expert-core are sufficiently large, our proposed algorithm gives the correct answer with high probability. 
Computational experiments using synthetic and real-world datasets 
demonstrated that our proposed answer aggregation algorithms outperform state-of-the-art algorithms. 

There are several directions for future research. 
Our model assumes that all experts give the correct answer with the same probability and all non-experts give an answer independently and uniformly at random. 
However, in reality, experts themselves may have different levels of expertise and non-experts may not be completely random. 
Although we have already confirmed that our proposed algorithms work well on real-world datasets, it is interesting to extend our model to such a more general scenario. 
Another direction is to generalize our model in a higher-level perspective. 
This study has focused on crowdsourced \emph{closed-ended} questions, where workers can select an answer from candidates. 
On the other hand, there are often cases where we wish to handle crowdsourced \emph{open-ended} questions, where workers have to answer without any candidates. 
We believe that our proposed algorithms may be applicable to this more general scenario by introducing a measure of similarity of answers (and thus similarity of workers). 

\section*{Acknowledgments}
This work was partially supported by JST ACT-I Grant Number JPMJPR17U79 and JSPS KAKENHI Grant Numbers 16K16005, 17H07357, 18J23034, and 19K20218.

\bibliographystyle{abbrv}
\bibliography{graph_mining_meets_crowdsourcing}
\section{Omitted proofs in Section~\ref{sec:theoretical}}\label{sec:omitted}
In this section, we give proofs for Theorems~\ref{thm:asymp_excore}--\ref{thm:mv_bad}.
To prove Theorems~\ref{thm:asymp_excore} and~\ref{thm:mv_good}, we use the Hoeffding's inequality.
\begin{theorem}[Hoeffding's inequality~\cite{Hoe1963}]
  Let $X_1,\dots,X_m$ be independent random variables such that $0\le X_i\le 1$ for all $i$.
  Let $X=\sum_{i=1}^m X_i/m$ and set $\mu=\mathbb{E}[X]$. Then, for all $\delta>0$, we have
\begin{align*}
\Pr\left[\left|X-\mu\right|\ge \delta\right]\le 2e^{-2m\delta^2}.
\end{align*}
\end{theorem}

\subsection{Proof of Theorem~\ref{thm:asymp_excore}}
\newtheorem*{thm:asymp_excore}{Theorem~2}
\begin{thm:asymp_excore}
  Let $W^*\subseteq \cW$ be the expert-core.
  If $\nex\ge 2$ and $m\ge \frac{2n^4\log\frac{n^2}{\epsilon}}{(\pex-1/s)^4}$ for $\epsilon>0$, then we have
  $\Pr[W^*\subseteq \cE]\ge 1-\epsilon$.
\end{thm:asymp_excore}
\begin{proof}
Let $r=\pex^2+(s-1)\cdot\left(\frac{1-\pex}{s-1}\right)^2~(=\frac{s}{s-1}(\pex-1/s)^2+1/s)$.
  Without loss of generality, we may assume $n>\nex\ge 2$ and $\epsilon\le 1$.
  We use the union bound to bound the probability of bad events.
  Let $\delta=\frac{(\pex-1/s)^2}{n^2}$ and $\rho=(\binom{n}{2}-\binom{\nex}{2})\cdot\frac{1}{s}+\binom{\nex}{2}\cdot r$.
  It should be noted that $1/s+2\delta<\rho/\binom{n}{2}<r-2\delta$ because
  \begin{align*}
    r-\frac{\rho}{\binom{n}{2}}
    &=\frac{\binom{n}{2}-\binom{\nex}{2}}{\binom{n}{2}}\cdot\left(r-1/s\right)
    =\frac{\binom{n}{2}-\binom{\nex}{2}}{\binom{n}{2}}\cdot\frac{s}{s-1}\left(\pex-1/s\right)^2
    > \frac{(\pex-1/s)^2}{\binom{n}{2}}>2\delta
  \end{align*}
  and
  \begin{align*}
    \frac{\rho}{\textstyle\binom{n}{2}}-1/s
    =\frac{\nex^2}{\binom{n}{2}}\cdot\left(r-1/s\right)
    =\frac{\nex^2}{\binom{n}{2}}\cdot\frac{s}{s-1}\left(\pex-1/s\right)^2
    >\frac{4(\pex-1/s)^2}{\binom{n}{2}}\ge 2\delta.
  \end{align*}
  We define some events for $u,v\in\cW$ as follows:
  \begin{description}
  \item[(a)] $E_1$: $1/s+\delta< p < r-\delta$;
  \item[(b)] $E_2(u,v)$: $m(r-\delta/2)\le \tau(u,v)\le m(r+\delta/2)$;
  \item[(c)] $E_3(u,v)$: $\tau(u,v)\le m(1/s+\delta/2)$.
  \end{description}

  Suppose that the events $E_1$, $E_2(u,v)$ occur for all $(u,v)\in \binom{\cE}{2}$, and 
  $E_3(u,v)$ for all $(u,v)\in \binom{\cW}{2}\setminus\binom{\cE}{2}$.
  Then, for $(u,v)\in \binom{\cE}{2}$, we have
  \begin{align*}
    \gamma(u,v)
    &\ge -\log\sum_{i=\lceil m(r-\frac{\delta}{2})\rceil}^m\binom{m}{i}(r-\delta)^i(1-r+\delta)^{m-i}
    \ge -\log(2e^{-\delta^2 m/2})
  \end{align*}
  by the Hoeffding's inequality.
  Also, for $(u,v)\in \binom{\cW}{2}\setminus\binom{\cE}{2}$, we have
  \begin{align*}
    \gamma(u,v)
    &\le -\log\!\!\!\!\sum_{i=\lceil m(\frac{1}{s}+\frac{\delta}{2})\rceil}^m\!\!\!\binom{m}{i}(\tfrac{1}{s}+\delta)^i(1-\tfrac{1}{s}-\delta)^{m-i}\\
    &=   -\log\left(1-\!\!\!\!\sum_{i=0}^{\lceil m(\frac{1}{s}+\frac{\delta}{2})\rceil-1}\!\!\!\binom{m}{i}(\tfrac{1}{s}+\delta)^i(1-\tfrac{1}{s}-\delta)^{m-i}\right)\\
    &\le -\log\left(1-2e^{-\delta^2 m/2}\right)
  \end{align*}
  by the Hoeffding's inequality.
  Since $m=\frac{4\log(n^2/\epsilon)}{\delta^2}\ge\frac{2\log(2n)}{\delta^2}$, we have
  \begin{align*}
    -(n-1)\log\left(1-2e^{-\delta^2 m/2}\right)
    &\le -\log\left(1-2(n-1)e^{-\delta^2 m/2}\right)\\
    &=-\log\left(2e^{-\delta^2 m/2}+1-2ne^{-\delta^2 m/2}\right)\\
    &\le -\log\left(2e^{-\delta^2 m/2}\right).
  \end{align*}
  Consider the sets $S_1,\dots,S_n$ computed in Algorithm~\ref{alg:peeling}.
  We have $d_{S_i}(v)\le -(n-1)\log\left(1-2e^{-\delta^2 m/2}\right)$ if $v\in S_i\setminus\cE$
  and $d_{S_i}(v)\ge -\log\left(2e^{-\delta^2 m/2}\right)$ if $v\in S_i\cap\cE$ and $|S_i\cap\cE|\ge 2$.
  Hence, $W^*\subseteq\cE$ holds. 

  The proof is completed by showing that
  the events $E_1$, $E_2(u,v)$ for $(u,v)\in \binom{\cE}{2}$, and 
  $E_3(u,v)$ for $(u,v)\in \binom{\cW}{2}\setminus\binom{\cE}{2}$
  occur at the same time with probability at least $1-\epsilon$.
  
  For each $q\in\cQ$,
  let $\kappa_q=\left|\{(u,v)\in \binom{\cW}{2}\mid l_{uq}=l_{vq}\}\right|$.
  Then we have $\mathbb{E}\left[\kappa_q\right]=\rho$.
  By the Hoeffding's inequality, we obtain
  \begin{align*}
    \Pr[\overline{E_1}]\le \Pr\left[\left|\tfrac{\sum_{q\in\cQ}\kappa_q}{m}-\rho\right|\ge \delta\cdot \binom{n}{2}\right]\le 2e^{-2m\delta^2}.
  \end{align*}
  Also, for $\{u,v\}\in\binom{n}{2}$, we get 
  \(\Pr[\overline{E_2}]\le 1-\Pr\left[\left|\tfrac{\tau(u,v)}{m}-r\right|\ge \tfrac{\delta}{2}\right]
  \le 2e^{-\frac{m\delta^2}{2}}\) by the Hoeffding's inequality since $\Pr[l_{uq}=l_{vq}]=r$.
  Similarly, for each non-expert $u\in\cW\setminus\cE$ and worker $v\in\cW\setminus\{u\}$,
  we have \(\Pr[\overline{E_3}]\le \Pr\left[\left|\tfrac{\tau(u,v)}{m}-r\right|\ge \tfrac{\delta}{2}\right]
  \le 2e^{-\frac{m\delta^2}{2}}\) by the Hoeffding's inequality since $\Pr[l_{uq}=l_{vq}]=1/s$.

  By the above claims, we have
  \begin{align*}
    \Pr\left[E_1\wedge\bigwedge_{(u,v)\in \binom{\cE}{2}}E_2(u,v)\wedge \bigwedge_{u\in\cW\setminus\cE,~v\in\cW\setminus\{u\}}E_3(u,v)\right]
    &\ge 1-\Pr[\overline{E_1}]
    -\!\!\!\!\!\!\!\!\sum_{(u,v)\in \binom{\cE}{2}}\!\!\!\!\Pr[\overline{E_2(u,v)}]
    -\!\!\!\!\!\!\!\!\!\!\!\!\sum_{(u,v)\in\binom{\cW}{2}\setminus\binom{\cE}{2}}
    \!\!\!\!\!\!\!\!\!\!\Pr[\overline{E_3(u,v)}]\\
    &\le 1-2e^{-2m\delta^2}-n(n-1)e^{-m\delta^2/2}\\
    &\le 1-n^2\cdot e^{-m\delta^2/2}
    \le 1-\epsilon. \qedhere
  \end{align*}  
\end{proof}

\subsection{Proof of Theorem~\ref{thm:mv_good}}
\newtheorem*{thm:mv_good}{Theorem~\ref{thm:mv_good}}
\begin{thm:mv_good}
  If $n=\nex\ge \frac{2\log\frac{2s}{\epsilon}}{(\pex-1/s)^2}$ for $\epsilon>0$, then we have
  $\Pr\left[\MV_q=a_q\right]\ge 1-\epsilon$ $(\forall q\in\cQ)$.
\end{thm:mv_good}
\begin{proof}
  Fix $q\in\cQ$ and let $\rho=\frac{\pex+1/s}{2}$.
  For each $c\in\cC$, let $I_c$ be an event such that $|u\in\cW\mid l_{uq}=c\}|\ge \rho\cdot n$.
  Note that, $\MV_q=a_q$ holds if $I_{a_q}\wedge \bigwedge_{c\in\cC\setminus\{a_q\}} \overline{I_{c}}$ occurs.
  
  By the Hoeffding's inequality, we obtain
  \begin{align*}
    \Pr\left[I_{a_q}\wedge \bigwedge_{c\in\cC\setminus\{a_q\}} \overline{I_{c}}\right]
    &=1-\Pr[\overline{I_{a_q}}]-\sum_{c\in\cC\setminus\{a_q\}}\Pr[I_c]\\
    &\ge 1- 2e^{-2n\cdot \left(\frac{\pex-1/s}{2}\right)^2}
    -2(s-1)e^{-2n\cdot \left(\frac{\pex-1/s}{2}\right)^2}\\
    &=1-2se^{-2n\cdot \left(\frac{\pex-1/s}{2}\right)^2}
    \ge 1-\epsilon.
  \end{align*}
  Hence, the claim holds.
\end{proof}

\subsection{Proof of Theorem~\ref{thm:mv_bad}}
\newtheorem*{thm:mv_bad}{Theorem~\ref{thm:mv_bad}}
\begin{thm:mv_bad}
  If $\nnon\ge \frac{\nex^2}{2\pi\epsilon^2}$ for $\epsilon>0$, then we have
  $\Pr\left[\MV_q=a_q\right]\le \frac{1}{2}+\epsilon$ $(\forall q\in\cQ)$.
\end{thm:mv_bad}
\begin{proof}
  Note that we have $\binom{2k}{k}=(2k!)/(k!)^2\le 4^k/\sqrt{\pi k}$ by Stirling's formula.
  Thus, we obtain
  \begin{align*}
    \Pr\left[\MV_q=a_q\right]
    &\le \sum_{i=n/2-\nex}^{\nnon}\binom{\nnon}{i} \left(\frac{1}{2}\right)^{\nnon}\\
    &\le \frac{1}{2}+\sum_{i=n/2-\nex}^{\nnon/2}\binom{\nnon}{i} \left(\frac{1}{2}\right)^{\nnon}
    \le \frac{1}{2}+\frac{\nex}{2}\cdot \binom{\nnon}{\nnon/2} \left(\frac{1}{2}\right)^{\nnon}\\
    &\le \frac{1}{2}+\frac{\nex}{2}\cdot \frac{2^{\nnon}}{\sqrt{\pi\nnon/2}} \left(\frac{1}{2}\right)^{\nnon}
    = \frac{1}{2}+\frac{\nex}{\sqrt{2\pi\nnon}}
    \le \frac{1}{2}+\epsilon.\qedhere
  \end{align*}  
\end{proof}

\end{document}